\documentclass[12pt]{article}

\usepackage[utf8]{inputenc}
\usepackage{multicol}
\usepackage{charter}
\usepackage{fullpage}
\usepackage[pdftex,bookmarks,hidelinks]{hyperref}
\usepackage{amssymb}
\usepackage{graphicx}
\usepackage{amsmath}
\usepackage{cleveref}
\usepackage[range-units=brackets, range-phrase=-]{siunitx}[=v2]
\usepackage[svgnames,dvipsnames,x11names,table]{xcolor}
\usepackage{lipsum}
\usepackage[sort&compress,numbers]{natbib}
\usepackage{hyperref}
\usepackage[total={6.5in,9in},top=1in,headsep=0.1in,headheight=1in]{geometry}

\usepackage{framed}

\makeatletter
\g@addto@macro\bfseries{\boldmath}
\makeatother

\newcommand\snowmass{
\begin{center}
  \rule[-0.2in]{\hsize}{0.01in}\\
  \rule{\hsize}{0.01in}\\
  \vskip 0.1in
  Submitted to the Proceedings of the US Community Study\\ 
  on the Future of Particle Physics (Snowmass 2021)\\
  \rule{\hsize}{0.01in}\\
  \rule[+0.2in]{\hsize}{0.01in}\\[-2em]
\end{center}
}

\usepackage[firstpage=true]{background}
\backgroundsetup{contents={\parbox{6.5in}{\snowmass}}, scale=1,placement=top,opacity=1,color=black,position={3.25in,1.2in}}

\usepackage{fancyhdr}
\fancypagestyle{plain}{%
  \fancyhf{}%
  \fancyhead[C]{}
  \fancyfoot[C]{\thepage}
}

\fancypagestyle{empty}{%
  \fancyhf{}%
  \fancyhead[C]{{\it Naturalness}}
  \fancyfoot[C]{\thepage}
}
\title{Naturalness: A Snowmass White Paper}
\date{}

\usepackage{authblk}

\author{Nathaniel Craig}

\medskip

\affil{Department of Physics, University of California, Santa Barbara, CA 93106, USA}

\begin{document}
\maketitle

 \begin{abstract}
\noindent We assess the state of naturalness in high-energy physics and summarize recent approaches to the three major naturalness problems: the cosmological constant problem, the electroweak hierarchy problem, and the strong CP problem.
\end{abstract}

\section{Introduction}

Writing about naturalness in the current era is fraught with peril, and with good reason. Of the three major naturalness problems in high-energy physics --- the cosmological constant problem, the electroweak hierarchy problem, and the strong CP problem --- it is difficult to reconcile natural solutions of the first with our understanding of physics at the eV scale, while natural solutions of the second are under intense pressure from the LHC's exploration of physics at the TeV scale. The third problem is in somewhat better condition, with decisive experimental tests still in the future, but its prevailing natural solutions face theoretical challenges of their own. In surveying this state of affairs, it is hard not to be pessimistic about the future prospects of naturalness-based reasoning.\footnote{There is another sense in which writing about naturalness is fraught with peril: there is already a great deal of excellent writing on the general subject. These include (among many others) the classic essay by Philip Nelson \cite{nelson}; Gian Giudice's 2008 \cite{Giudice:2008bi} and 2013 \cite{Giudice:2013yca} essays on naturalness and the hierarchy problem; Hitoshi Murayama's ICTP summer school lectures on supersymmetry \cite{Murayama:2000dw}; Markus Luty's TASI lectures on supersymmetry breaking \cite{Luty:2005sn}; Steve Martin's supersymmetry primer \cite{Martin:1997ns} and his forthcoming book with Herbie Dreiner and Howie Haber; James Wells' articles on fine-tuning and naturalness \cite{Wells:2013tta, Wells:2016luz, Wells:2018sus, Wells:2018yyb, Wells:2021zdp}; Tim Cohen's TASI lectures on effective field theory \cite{Cohen:2019wxr} and Cliff Burgess' textbook on the same topic \cite{Burgess:2020tbq}; Michael Dine's \cite{Dine:2000cj} and Anson Hook's \cite{Hook:2018dlk} TASI lecture notes on the strong CP problem; Matthew McCullough's TRISEP lecture notes on the hierarchy and strong CP problems \cite{McCullough:2018knz}; Steve Weinberg's review of the cosmological constant problem \cite{Weinberg:1988cp}; Joe Polchinski's Solvay lecture \cite{Polchinski:2006gy}, Raphael Bousso's TASI lectures \cite{Bousso:2007gp}, and Cliff Burgess' Les Houches lectures \cite{Burgess:2013ara} on the same topic; and Seth Koren's Sakurai Dissertation Prize-winning thesis on the hierarchy problem \cite{Koren:2020biu}. This white paper is also far from the only Snowmass contribution dedicated to aspects of naturalness; see e.g.~\cite{Blinov:2022tfy, Agrawal:2022yvu, Berglund:2022qcc, Draper:2022pvk, Batell:2022pzc, Dvorkin:2022jyg, Asadi:2022njl}, many of which go into much greater detail about specific approaches to naturalness problems than this paper.}  

But it bears remembering that these three naturalness problems are the three {\it outstanding} naturalness problems in high-energy theory --- the ones whose solutions remain unknown. So many measured quantities in the Standard Model are consistent with naturalness-based reasoning, and in notable cases were {\it predicted} by it. When focusing on the open problems of naturalness, we are prone to forgetting its past solutions. In truth, the abundant past validation of naturalness is a compelling reason to take it seriously when confronting the problems at hand. 

This perspective --- that naturalness is a pragmatic and empirically-validated strategy for discovering new physics --- is far from the only rationale that has been used to support natural reasoning. Throughout its history, naturalness has been variously framed as a pragmatic strategy, a bedrock principle, an aesthetic criterion, and a catastrophic folly. In truth, it is a bit of each. Faced by an essentially infinite space of candidates for the theory of Nature and a very finite number of experimental tests (at least in one lifetime), physicists must come up with strategies to focus experimental efforts in directions that seem most likely to yield discovery. Occasionally these strategies emerge from considerations of simplicity, or admit codification into robust principles. Frequently, they do not. Ultimately, they are judged less by these considerations than by their success or failure in predicting new physics.

In this white paper, at least, we will focus primarily on naturalness as a pragmatic strategy for discovery. It is a strategy that succeeds again and again as we go up through the mass scales of Standard Model particles, making it reasonable to expect its continued success in addressing the problems at hand. Of course, to the extent that this rationale is inductive, we should be mindful of the well-known problems of induction. Bertrand Russell perhaps put it best \cite{russell}:
\begin{quote}
The man who has fed the chicken every day throughout its life at last wrings its neck instead, showing that more refined views as to the uniformity of Nature would have been useful to the chicken.
\end{quote}
The many successful postdictions and handful of successful predictions of naturalness do not guarantee its continued success. Perhaps the cosmological constant problem and electroweak hierarchy problem are signaling that more refined views as to the uniformity of Nature would have been useful to the particle theorist. 

But even that would be a useful outcome of natural reasoning. Whether it be the cosmological constant, the Higgs mass, or the theta angle of QCD, there are essentially two possibilities: either the parameter in question is natural, and we need to understand the mechanism; or the parameter is unnatural, and we need to develop the ``more refined views as to the uniformity of Nature.'' In the former case, we may already have come up with the essential mechanism and are simply in need of experimental direction to confirm it, or we may need to discover the mechanism in its entirety. In the latter case, the failure of naturalness is interesting in its own right: it signifies that Nature works in a way that is fundamentally different from what previous examples have led us to expect. 

Of course, this does not justify thinking about naturalness {\it ad infinitum}. Challenges to the naturalness strategy are both beneficial and necessary. They serve to sharpen our understanding of what naturalness is, and what it is not. And ultimately, if we have exhausted all promising paths, these challenges should persuade us to turn our undivided attention towards other strategies for discovery. 

But that day has not yet come. The three marquee naturalness problems are, as yet, undecided. In each case, some of the most appealing solutions have been experimentally tested and found wanting. From this we have learned something about the paths that Nature chose not to take, and been compelled to go off in search of paths less traveled. And what paths we have discovered! The past decade has seen a proliferation of new ideas about how the naturalness strategy might play out, and the coming decade will surely see even more. For the most part, the attendant experimental tests are close at hand, though often in places we had not yet thought to look. 

The primary goal of this white paper is to sketch the new paths for naturalness that have been discovered in the past decade, point to some of the possible paths that may be explored in the coming decade, and highlight the wealth of future experimental tests. To this end, we begin with a summary of historical reasoning behind naturalness as a pragmatic strategy, its codification in the language of technical naturalness and 't Hooft naturalness, and its susceptibility to anthropic reasoning. We then turn to the three naturalness problems of the era: the strong CP problem, the cosmological constant problem, and the electroweak hierarchy problem. When these three problems are discussed together, they are typically introduced in order of ascending or descending dimensionality of the operators involved. Here we will take an alternative approach, running from the theta angle of QCD, to the cosmological constant, to the mass of the Higgs, in the hope of drawing a clearer line through common solutions to the different problems. In each case we summarize the problem and historical approaches before turning to recent developments. We conclude by looking to the future, with an eye towards the promise of developments on the horizon.

\section{A Brief History of Naturalness}

Although our aim is to look forwards, rather than backwards, there are some instructive lessons to be learned from exploring the history of naturalness. What we would now recognize as naturalness arguments have a long history, dating back at least to Copernicus.\footnote{For a broad overview, see the classic essay by Philip Nelson \cite{nelson}.} Within the field of high-energy physics, definite notions of naturalness emerged in the 1930's. Perhaps the two most striking exemplars are Weisskopf's calculation of the self-energy of the electron \cite{Weisskopf:1934gig, Weisskopf:1939zz} and Dirac's Large Numbers Hypothesis \cite{Dirac:1938mt}. While both are examples of what we would now classify as naturalness-based reasoning, they are wildly different in both their nature and their effectiveness, and it is instructive to compare them.

Weisskopf began with the observation that the quantum theory of a relativistic electron with an ultraviolet cutoff $\Lambda$ suffers from a number of divergent contributions to the electron self-energy. The most striking is the contribution $\Delta m_e \sim e^2 \Lambda$, which is the familiar divergence of the classical Coulomb self-energy corresponding to an electron radius $r \sim 1/ \Lambda$. The problem is exacerbated by the introduction of quantum mechanics, as quantum fluctuations of the electromagnetic field around the electron contribute as $\Delta m_e \sim e^2 \Lambda^2 / m_e$. Of course, these divergences can all be absorbed in a proper renormalization procedure, which wouldn't be developed until some time after Weisskopf's calculation. But if one supposes that $\Lambda$ carries some physical significance, for $\Lambda \gg m_e$ this naively implies finely-tuned cancellations among large contributions $\Delta m_e$ to the electron self-energy in order to obtain the measured value of $m_e$. 

However, there is an alternative: additional degrees of freedom could appear to modify the calculation and avoid the need for fine-tuning. And indeed, this is precisely what Weisskopf found. As Dirac had proposed, a relativistic quantum theory of the electron requires a new degree of freedom, the positron (whose appearance can be attributed to a new symmetry of the quantum theory, CPT). The appearance of the positron modifies the calculation of the electron self-energy, and the power divergences cancel between the virtual contributions of electrons and positrons. All that remains is a logarithmic sensitivity, $\Delta m_e \sim \frac{3 e^2}{2 \pi} m_e \log (\Lambda/m_e),$ such that $\Delta m_e$ does not greatly exceed $m_e$ even for $\Lambda \gg m_e$. 

In his second paper on the subject \cite{Weisskopf:1939zz}, Weisskopf considered the analogous self-energy of a charged scalar. In this case there are no additional degrees of freedom that appear to cancel the divergences. Weisskopf then posited precisely the sort of naturalness argument that we would recognize today, taking the critical length $a \equiv \Lambda^{-1}$ to be set by the mass of the scalar itself \cite{Weisskopf:1939zz}:

\begin{quote}
This may indicate that a theory of particles obeying Bose statistics must involve new features at this critical length, or at energies corresponding to this length; whereas a theory of particles obeying the exclusion principle is probably consistent down to much smaller lengths or up to much higher energies.  
\end{quote}

Of course, Weisskopf's conclusions should be viewed through the subsequent lens of renormalization. Properly speaking, the divergences he encountered and parameterized with a cutoff $\Lambda$ can be absorbed by a suitable renormalization prescription. This is perfectly adequate if we only expect our theory to relate measurements in different channels and at different scales. Thus the perfectly sensible observation that the Standard Model {\it on its own} does not suffer from a hierarchy problem --- it simply has parameters which are fixed by measurements, and divergences encountered in loop calculations are duly absorbed by renormalization. But if we expect the fundamental theory to be finite and fully predictive, then Weisskopf's divergences take on a different character entirely. From this perspective, the divergences arise because we are only computing in a subset of the full theory, and their existence signals the approximate size of finite, physical contributions from the missing parts of the theory. 

Given the key role that Dirac played in Weisskopf's work, it is surprising that the conclusions Dirac drew from his own contemporary natural reasoning were quite far off the mark. Dirac's underlying expectation was that ``any two of the very large dimensionless numbers occurring in Nature are connected by a simple mathematical relation, in which the coefficients are of the order of magnitude unity'' \cite{Dirac:1938mt}. Although this expectation ultimately underlies more modern notions of naturalness, Dirac's inferences took their own path. Dirac understood that there was a mass scale associated with gravity, $M_{\rm Pl} \sim 10^{19}$ GeV, as well as a mass scale associated with the proton, $m_p \sim 1$ GeV, and wished to understand why $m_p \ll M_{\rm Pl}$. In Dirac's own framing, the goal was to explain why $\frac{G m_p^2}{\hbar c} \sim 5 \times 10^{-39}.$ Dirac noted that the Hubble age of the universe was about $T \frac{m_p c^2}{\hbar} \sim 10^{42}$, and that the mass of the universe to its visible limits was about $\frac{M}{m_p} \sim (10^{40})^2$. To him this suggested that there was a causal connection between dimensionless constants and powers of $T$. Since $T$ changes in time, that also implies that fundamental constants change in time, e.g., that $G$ evolves as $1/t$, and that $M$ evolves as $t^2$. He proceeded to develop an elaborate theory of cosmology around this idea, which Nature does not support. 

Although the true explanation for the proton mass eluded Dirac, we now understand it to be a beautiful triumph of naturalness criteria. The answer is that the proton mass is dynamically generated by confinement, which in turn arises from the logarithmic evolution of a dimensionless coupling, which itself is the manifestation of a violation of symmetry --- in this case, (classical) conformal symmetry. This phenomenon, {\it dimensional transmutation}, explains the existence of exponentially different scales. So Dirac's question was a good one, and the ultimate answer is a triumph of natural reasoning. The fact that Dirac's own answer was spectacularly incorrect carries an important lesson: {\it the failure of a specific answer to a naturalness problem does not signify the failure of the problem itself.}

In the meantime, Weisskopf's reasoning succeeds again and again as we proceed up in scale from the electron. Every fermion we encounter enjoys the same resolution to its self-energy puzzle as the electron. Although an apparently fundamental scalar would not appear until the discovery of the Higgs, the light pseudo-scalar bound states of the strong interactions provide a clear validation of Weisskopf's reasoning for scalars. The lightest of the charged mesons, the $\pi^\pm$, experience precisely the same divergent contribution to their self-energy that Weisskopf computed in 1939. As this is not shared by the neutral $\pi^0$, the divergent electromagnetic contribution to the self-energy can be framed in terms of the difference in the squared masses,
\begin{equation}
m_{\pi^\pm}^2 - m_{\pi^0}^2 =  \frac{3 \alpha}{4 \pi} \Lambda^2 \, .
\end{equation}

Given the size of the charged-neutral meson splittings, $m_{\pi^\pm}^2 - m_{\pi_0}^2 \sim (35.5 \, {\rm MeV})^2$, we expect the loop should be cut off around 850 MeV if electromagnetic loops explain the mass difference.  Lo and behold, the $\rho$ meson enters at 775 MeV, which provides a cutoff for the effective theory as the harbinger of compositeness. In fact, the argument can be made even more precise. Using Weinberg's sum rules and assuming the lightest vector ($\rho$) and axial vector ($a_1$) mesons dominate, the cutoff dependence is replaced by dependence on the masses of the vector and axial vector mesons \cite{Das:1967it}:
\begin{equation}
m_{\pi^\pm}^2 - m_{\pi^0}^2 \approx  \frac{3 \alpha}{4 \pi} \frac{m_\rho^2 m_{a_1}^2}{m_\rho^2 + m_{a_1}^2} \log \left( \frac{m_{a_1}^2}{m_\rho^2} \right) \, .
\end{equation}
Although the naive cutoff dependence in the low-energy theory of the pions alone is unphysical (depending on the choice of regularization, and entirely removable by renormalization), it nonetheless provides a useful proxy for the dependence on physical scales in a more complete theory.

Of course, both the electron and charged pion masses are post-dictions: in each case, the new features appearing at Weisskopf's ``critical length'' were already known, and the self-energy calculation was merely a validation that the known particles and interactions fit together in a natural way. But the same logic has also been used to make successful predictions, most notably the prediction of the charm quark mass by Gaillard and Lee in 1974 \cite{Gaillard:1974hs}. By then it was well known that mass difference between the $K_L^0$ and $K_S^0$ states in a theory with only the up, down, and strange quarks was quadratically divergent,
\begin{equation}
m_{K_L^0} - m_{K_S^0}= \simeq \frac{1}{16 \pi^2} m_K f_K^2 G_F^2 \sin^2 \theta_C \cos^2 \theta_C \times \Lambda^2 \, ,
\end{equation}
where $f_K = 114$ MeV is the kaon decay constant and $\sin \theta_C = 0.22$ is the Cabibbo angle. Requiring this correction to be smaller than the measured value $(M_{K_L^0} - M_{K_S^0} )/M_{K_L^0} = 7 \times 10^{-15}$ gives $\Lambda \lesssim 3$ GeV. Extending the theory to include the charm quark as proposed by Glashow, Iliopoulos, and Maiani \cite{Glashow:1970gm}, the divergence is eliminated and instead replaced by corresponding dependence on the mass of the charm quark. Once again, the unphysical divergences encountered in part of the theory anticipate finite contributions arising in the full theory. Gaillard and Lee's corresponding prediction $m_c \lesssim 1.5$ GeV presaged the discovery of the charm quark at $m_c \simeq 1.2$ GeV in the same year.

\subsection{In search of a principle}

Weisskopf's natural reasoning eventually gave way to the familiar narrative of Wilson \cite{Wilson:1970ag}, Weinberg \cite{Weinberg:1975gm}, Susskind \cite{Susskind:1978ms}, 't Hooft \cite{tHooft:1979rat}, and Veltman \cite{Veltman:1980mj}. An improved understanding of the structure of radiative corrections supported Dirac's original expectation of naturalness  --- that all dimensionless quantities should be order-one in the appropriate units --- while recognizing that selection rules could lead to more refined criteria.

In particular, parameters have come to be known as {\it technically natural} if their size in the ultraviolet theory is not spoiled on the way to the infrared by physics at intermediate scales. Technical naturalness may be assured by approximate symmetries: certain symmetries can control the form of quantum corrections, and when these symmetries are broken the quantum corrections must be proportional to the symmetry-breaking itself. This led to the somewhat narrower notion of {\it 't Hooft naturalness}: a parameter is {\it 't Hooft natural} if symmetries are restored when the parameter is set to zero. Most observed hierarchies are both technically natural and 't Hooft natural, but it is possible for parameters to be technically natural without being 't Hooft natural. The latter is sufficient to assure the former, but not necessary. In cases where a parameter is technically natural without being 't Hooft natural, an assumption is typically made about the absence of additional physical thresholds intermediate between the ultraviolet and the infrared, which could induce large corrections in the absence of a symmetry.

These refinements help to clarify what sort of small numbers pose naturalness problems, and what do not. Circling back to Weisskopf, the electron mass is 't Hooft natural due to the chiral symmetry restored in the $m_e \rightarrow 0$ limit, while the mass of a charged scalar is not. More broadly, all the fermionic mass hierarchies in the Standard Model are 't Hooft natural, stemming from the violation of Standard Model flavor symmetries that are restored when Yukawa couplings are taken to zero. It is still worth asking how these observed hierarchies came about, but whatever mechanism explains the flavor hierarchy could live in the far ultraviolet, given that its predictions would persist undisturbed into the infrared. In contrast, the cosmological constant, Higgs mass, and QCD theta angle are neither technically nor 't Hooft natural in generic extensions of the Standard Model.

\subsection{An alternative to naturalness}

How could the logic of naturalness fail? Amusingly, Dirac's own attempts at natural reasoning led to the identification of a possible failure mode. Responding to Dirac, in 1961 Dicke pointed out that questions about the age of the universe could only arise if conditions were right for the existence of life, with the specific criteria that the universe must be old enough so that some stars completed their time on the main sequence and produced heavy elements, and young enough that some stars were still undergoing fusion \cite{1961Natur.192..440D}. Working these out in terms of fundamental units, Dicke found the upper and lower bounds essentially lead to Dirac's relations --- but rather than resulting from time variation of fundamental parameters, they followed entirely from the existence of observers. 

Dicke's argument is perhaps the first use of what has subsequently been termed {\it anthropic reasoning} in modern physics \cite{Weinberg:1988cp}, though the term itself would not be coined for another twelve years. At heart, anthropic reasoning stems from the fact that observed parameters are necessarily compatible with the existence of an observer. To gain {\it explanatory} power over the values of fundamental parameters, it requires something like the existence of alternate universes in which these fundamental parameters vary, as well as some assumptions about the distribution of parameters among these universes. If the most natural values of certain parameters do not lead to the formation of suitable observers, then anthropic reasoning in this context allows us to understand why we might instead observe a universe with unnatural values.

Of course, the possible role of anthropic reasoning in circumventing a naturalness problem depends sensitively on the parameter in question, and the extent to which its variation can be tied to the formation of observers. To say more requires us to commit to the specifics of a problem. 

\section{The Strong CP Problem}

We know that CP is not a symmetry of the Standard Model, being broken by the weak interactions. But there is another potential source of CP violation that is not, as yet, observed. The QCD Lagrangian in principle contains a term of the form
\begin{equation}
\mathcal{L} \supset  - \theta \frac{\alpha_s}{8 \pi} G^a_{\mu \nu} \widetilde G^{a \mu \nu} \, .
\end{equation}
The $\theta$ term is $P$- and $T$-odd, hence $CP$-odd. While this can be written as a total derivative, for non-abelian theories we are not entitled to discard boundary terms due to the existence of instantons, and so we have to contend with the possible physical consequences.
 
The effects of the $\theta$ term in the QCD Lagrangian can be traced down to low energies, where they contribute to a host of hadronic CP-violating observables. For instance, in the pion-nucleon effective Lagrangian it induces a CP-violating nucleon-nucleon-pion coupling. At one loop this yields a contribution to the neutron electric dipole moment of order
\begin{align}
d_n  \sim 5 \times 10^{-16} \, \theta \, e \,{\rm cm} \, ,
\end{align}
whereas the experimental bound is $|d_n| < 1.8 \times 10^{-26} \, e$ cm; an inferred bound from the $^{199}$Hg EDM limit is comparable. This implies $\theta \lesssim 10^{-10}$. Given that CP is not a symmetry of the Standard Model, the natural expectation might have been $\theta \sim \mathcal{O}(1)$, amounting to a violation of naturalness expectations by ten orders of magnitude. This is known as the strong CP problem. See e.g.~\cite{Dine:2000cj, Peccei:2006as, Hook:2018dlk} for excellent overviews.

It bears emphasizing that $\theta$ is technically natural (by the definition used here) {\it when restricted solely to the Standard Model \cite{Ellis:1978hq}}. CP violation from the CKM matrix only feeds in at high loop order, generating a contribution to $\theta$ that is well below current limits. However, it is not 't Hooft natural, as CP symmetry is not restored in the Standard Model when $\theta \rightarrow 0$. It is not obvious why $\theta$ should be small in the UV, but even if it were, its smallness would generically be spoiled by physics beyond the Standard Model at intermediate scales.

There are three conventional avenues for rendering $\theta$ natural. The first is to have a massless quark, since then $\theta$ is unphysical as it may be removed entirely by redefinitions of the massless quark. The second is to solve the problem in the UV by imposing P or CP as exact symmetries at a high scale, broken spontaneously to induce the CP violation inherent to the CKM phase without generating large corrections to $\theta$. The third is to solve the problem in the IR by relaxing the value of $\theta$. The first option is ruled out by lattice data, which strongly disfavors a massless quark.\footnote{For an excellent and very recent summary of the state of affairs, see \cite{Davies:2022ueb}. For a novel model leveraging a massless quark in a hidden sector to control the $\theta$ angle of the Standard Model, see \cite{Hook:2014cda}.} Instead, let us briefly explore the second and third options.

\subsection{UV Solutions}

Perhaps the most transparent option is to render the theta angle small by reference to the ultraviolet: make P or CP good symmetries in the UV, broken spontaneously at some scale to give the known CP violation observed in the CKM matrix. Recent progress has been surveyed in a dedicated Snowmass white paper \cite{Blinov:2022tfy}, and so here we will restrict ourselves to sketching the key ideas and developments.

The physical strong CP angle is the combination of the quark mass term phase and the intrinsic QCD phase,
\begin{equation}
\bar \theta = \theta + {\rm arg} \det M = \theta + {\rm arg} \det [Y_u Y_d] \, .
\end{equation}
The challenge for a technically natural approach is thus to explain why  ${\rm arg} \det [Y_u Y_d]$
 is small, but the combination that picks out the phase in the CKM matrix, ${\rm arg} \det [Y_u Y_d - Y_d Y_u],$
 is not. This entails no small degree of cleverness. The most common route, the Nelson-Barr mechanism (\cite{Nelson:1983zb} and \cite{Barr:1984qx}), starts with CP as a UV symmetry and breaks it via the vevs of some complex scalars, which accumulate a relative phase. These scalars couple to Standard Model quarks with the assistance of additional vector-like quarks, and couplings are engineered (with the assistance of exact $\mathbb{Z}_2$ global symmetries) in such a way as to guarantee that ${\rm arg} \det [Y_u Y_d] = 0$ at tree level but the CKM phase is nonzero. Much of the complexity of Nelson-Barr models stems from the fact that they protect $\theta$ with a symmetry (CP) that the Standard Model breaks in a non-decoupling manner.
 
 An alternative avenue is to leverage the second symmetry that can protect $\theta$, namely parity. Although parity is also violated by the Standard Model, a generalized parity $P$ may be restored in the UV by e.g.~extending $SU(2)_L \rightarrow SU(2)_L \times SU(2)_R$ and having parity act as conventional parity supplemented by $P: SU(2)_L \leftrightarrow SU(2)_R$. The $\theta$ term is odd under this parity, and so is forbidden in the UV where the parity is good. Unlike Nelson-Barr models, parity violation is not needed to allow for a complex CKM phase; it's already allowed, and simply mirrored by a parity phase for the mirror fields. Of course, this parity must be spontaneously broken in order to satisfy direct limits on $SU(2)_R$ gauge bosons and other states. 
 
Additional challenges arise in both cases. The success of UV models is contingent upon their specific field content, and can easily be spoiled by additional fields and generic interactions; this is particularly challenging when extending these models to address other problems such as the electroweak hierarchy problem. Although UV models are not as susceptible to quality problems as their IR competitors (about which more will be said momentarily), the requisite scale of P or CP violation is low enough to introduce questions about the origin and stability of large parametric hierarchies. These challenges are comprehensively summarized in \cite{Dine:2015jga, Albaid:2015axa}.
 
Challenges aside, UV approaches to the strong CP problem are a promising and relatively unexplored direction, as they point to a wide array of experimental signatures not traditionally associated with strong CP. They have correspondingly seen something of a revival in recent years; developments include 
\begin{itemize}
\item A new approach to spontaneous parity breaking associated with a vanishing Higgs quartic at high energies \cite{Hall:2018let}. 
\item Renewed focus on parity-based solutions and their collider signatures, either with \cite{Craig:2020bnv} or without \cite{DAgnolo:2015uqq} large vector-like mass terms for $SU(2)$-singlet fermions. 
\item Sharpened predictions for two-loop contributions to $\theta$ in minimal parity models \cite{deVries:2021pzl} and Nelson-Barr models \cite{Valenti:2021rdu}, suggesting near-future tests of UV models for strong CP with hadronic EDMs.
\item The development of a comprehensive framework for the UV completion of Nelson-Barr models without numerical coincidences or small dimensionless input parameters \cite{Valenti:2021xjp}.
\end{itemize}

The space of UV solutions to strong CP is far from being fully explored, and the promise of near-future experimental tests (at both colliders and EDM experiments, as well as unexpected venues such as gravitational wave detectors \cite{Craig:2020bnv}) is highly compelling. We refer the reader to  \cite{Blinov:2022tfy} for a discussion of some promising opportunities.

\subsection{IR Solutions}

If parity and Nelson-Barr models can be said to solve the strong CP problem in the UV, the axion solves the strong CP problem in the IR. Recent progress and outstanding questions in axion theory are comprehensively summarized by a dedicated Snowmass white paper \cite{Agrawal:2022yvu}, and here we provide only a cursory overview with an eye towards naturalness-related considerations.

The basic idea is simple: if we can introduce a pseudoscalar field $a$ that couples to $G \tilde G$ like $\theta$, i.e., 
\begin{equation}
\Delta \mathcal{L} = - \frac{\alpha_s}{8\pi} \left(\theta + \frac{a}{f_a} \right) G \tilde G
\end{equation}
then the total effective CP violating angle is $\theta + \langle a \rangle / f_a$ and the QCD vacuum energy becomes
\begin{equation}
E(a,\theta) = - m_\pi^2 f_\pi^2 \cos (\theta + a/f_a) \, .
\end{equation}
This has a minimum at $\langle a \rangle = - \theta f_a$ where the total effective CP violating angle is set to zero, solving the strong CP problem. 

The requisite coupling of the pseudoscalar emerges automatically if it is the goldstone boson of a spontaneously broken symmetry, the Peccei-Quinn symmetry $U(1)_{PQ}$ \cite{Peccei:1977hh}. 
This doesn't work in the Standard Model with one Higgs doublet; both the Higgs and its conjugate are involved in Yukawa couplings, so there is no way of assigning $U(1)_{PQ}$ charges to the Higgs and quarks such that Yukawa terms are invariant. Introducing a second doublet leads to a highly successful and predictive framework, the Weinberg-Wilczek model \cite{Weinberg:1977ma, Wilczek:1977pj} in which the axion decay constant is $1/v$, which is ruled out by direct searches. This idea can be easily rescued by adding a singlet complex scalar also transforming under $U(1)_{PQ}$ to obtain the DFSZ (Dine-Fischler-Srednicki-Zhitnisky) axion \cite{Zhitnitsky:1980tq, Dine:1981rt}. If the singlet acquires a much larger vev, this renders the axion lighter and more weakly coupled. Another option is to introduce new fermions charged under QCD; this gives the KSVZ (Kim-Shifman-Vainstein-Zakharov) axion \cite{Kim:1979if, Shifman:1979if}. Although both DFSZ and KSVZ axions make relatively sharp predictions for axion couplings to Standard Model particles, the interplay between multiple axions (via e.g. the clockwork mechanism \cite{Kaplan:2015fuy}) can open entirely new parameter space \cite{Farina:2016tgd}.

Axion solutions to strong CP are compelling for a number of reasons. Axions generically arise in string compactifications, making their appearance in association with the strong CP problem less surprising. Famously, they can also furnish a dark matter candidate with a predictive mechanism for the dark matter relic abundance. Unlike UV solutions to strong CP, they are insensitive to new sources of CP violation from additional degrees of freedom beyond the Standard Model; axions ultimately relax the sum of all contributions to $\theta$. However, this requires that the $U(1)_{PQ}$ symmetry giving rise to the axion be exceptionally good (with the {\it singular exception} of being anomalous with respect to QCD), much better than generic expectations about the violation of global symmetries in a theory of quantum gravity would suggest. This axion quality problem \cite{Barr:1992qq, Kamionkowski:1992mf, Holman:1992us} has long motivated model-building solutions \cite{Randall:1992ut}, with a number of new approaches emerging in recent years  \cite{Lillard:2018fdt, Cox:2019rro, Ardu:2020qmo, Heidenreich:2020pkc, Darme:2021cxx}. Recent investigation suggests that the axion quality problem may be ameliorated in weakly-coupled string compactifications, as large hierarchies among instanton actions lead to exponential suppression of dangerous PQ-violating effects \cite{Demirtas:2021gsq}. It is unclear whether this should persist for compactifications outside the weakly-coupled regime.

As with UV solutions to strong CP, the space of axion solutions to the strong CP problem is far from comprehensively explored, particularly in relation to the axion quality problem. We refer the reader to \cite{Agrawal:2022yvu} for a discussion of some of the most promising opportunities. 

\subsection{Anthropics}

Rather than being accidental, or rendered natural by one of the mechanisms discussed above, the strong CP problem could be explained with anthropic reasoning if values of $\theta$ much larger than the true value were unfavorable to observers. This is not a typical line of reasoning for the strong CP problem since it is not obvious if any of the important processes in the early universe are significantly altered by variation of $\theta$ provided $\theta \lesssim 0.1$ \cite{Lee:2020tmi}. However, if one assumes that the cosmological constant is anthropically determined (about which more momentarily), by making sufficiently strong assumptions about the mechanism one can obtain a correlated bound on $\theta$ \cite{Kaloper:2017fsa}. As with many anthropic arguments, the devil is in the details \cite{Dine:2018glh}. Of course, there is nothing wrong with the coexistence of anthropic explanations for the smallness of some parameters and natural explanations for others. Perhaps anthropic explanations for strong CP will seem more compelling if decisive experimental tests of natural explanations come up empty. We are far from this situation at present, and the strong CP problem remains a compelling target for natural reasoning.

\section{The Cosmological Constant Problem}

Let us now move to the other end of the dimensional spectrum, to the cosmological constant. There are a number of excellent overviews of the cosmological constant problem, e.g. \cite{Weinberg:1988cp, Carroll:2000fy, Padmanabhan:2002ji, Nobbenhuis:2004wn, Polchinski:2006gy, Bousso:2007gp}, and we shall not attempt to improve on them here. 

Conventionally, the cosmological constant $\Lambda$ (not to be confused with the momentum cutoff $\Lambda$ appearing elsewhere) is a constant term that can appear in the Lagrangian of a theory of gravity. This corresponds to a vacuum energy density $\rho_\Lambda \equiv \frac{\Lambda}{8 \pi G}$; as the former quantity is the one most relevant to both experimental measurements and the discussion of quantum corrections, we'll focus on $\rho_\Lambda$ in what follows, and revert to using $\Lambda$ as a momentum cutoff.  

Decisive evidence for nonzero $\rho_\Lambda$ was accumulated in 1998 from observations of distance-redshift relations for Type 1a supernovae, and further solidified by CMB measurements yielding a preferred value of
\begin{equation}
\rho_\Lambda = (2.26 \times 10^{-3}   \, {\rm eV})^4 \, .
\end{equation}
Purely on dimensional grounds, in a field theory with a cutoff $\Lambda$ coupled to gravity we expect $\rho_{\Lambda} \propto \Lambda^4$. Radiatively, vacuum bubbles of a field of mass $m$ in an effective field theory with cutoff $\Lambda$ contribute 
\begin{equation}
\rho_{\Lambda} \simeq c_0 \Lambda^4 +  c_2 m^2 \Lambda^2 +  c_4 m^4 + \dots
\end{equation}
where the coefficients $c_i$ are on the order of $1/16 \pi^2$ in natural units. In the Standard Model, if we take $\Lambda \sim M_{\rm Pl}$, then we expect $\rho_\Lambda \sim 10^{120} \rho_{\Lambda, \, {\rm obs}},$
 an enormous violation of naturalness expectations. This is the cosmological constant problem. 

There are two points to emphasize about the underlying problem. The first is the cutoff dependence; as we have discussed, the cutoff itself is unphysical, but gives us a plausible proxy for finite contributions associated with new degrees of freedom at the scale $\Lambda.$  However, it bears emphasizing that there is an enormous problem even if we discard the power-law cutoff dependence and keep only the finite and log-dependent terms, which contribute at $\mathcal{O}(m^4)$. In the Standard Model, the finite contribution from the top quark already implies $\rho_{\Lambda} \sim 10^{53} \rho_{\Lambda, \, {\rm obs}}.$ The cosmological constant is far from technically natural even when restricted solely to the Standard Model. 

The second consideration is whether we somehow misunderstand how QFT couples to gravity --- perhaps the estimate of quantum contributions to the cosmological constant is flawed on the grounds that we misunderstand how gravity couples to virtual particles. But we know that a virtual electron contributes to the vacuum polarization correction to the Lamb shift, and by the equivalence principle this must couple to gravity. Loops are ``real'' in terms of their observable consequences. And although we have focused on quantum corrections to $\rho_\Lambda$, there is also a problem at tree level, as phase transitions induce changes in the vacuum energy density. We should expect a contribution of order $\rho_\Lambda \sim \Lambda_{QCD}^4$ from the QCD phase transition alone, much less contributions from the electroweak phase transition or potential earlier phase transitions associated with unification or sectors other than our own.

Numerous solutions to the cosmological constant problem have been proposed; for an extremely comprehensive enumeration of possibilities prior to 2004, see \cite{Nobbenhuis:2004wn}. In what follows we'll review a subset of these approaches with an eye towards recent progress.

\subsection{Anthropics}

The explanation for the cosmological constant that is both most popular and most controversial among high-energy physicists (at present) is the one that discards naturalness in favor of anthropic reasoning. For observers to be present in order to {\it see} a universe with a small cosmological constant, the cosmological constant must be small enough that sufficiently large gravitationally bound systems can form. By sufficiently large, we have in mind something that forms stars and planets, which requires heavy elements --- so the structures of interest are galaxies or globular clusters. 

The anthropic argument for the cosmological constant is often credited to Weinberg \cite{Weinberg:1987dv}, and with good reason. But it also bears emphasizing that a general sketch of the argument was made by Banks in 1985 \cite{Banks:1984cw}, and a qualitative bound along the lines of Weinberg's was made by Barrow \& Tipler in 1986 \cite{Barrow:1988yia}. In any event, a simplified version of Weinberg's argument will suffice for our purposes. We know that in our universe gravitational condensation had already begun at redshift $z_c \geq 4$ (from the redshifts of the oldest quasars), when the energy density was greater than the present mass density $\rho_{M_0}$ by a factor $(1+z_c)^3$. A cosmological constant has little effect as long as the non-vacuum energy density is larger than $\rho_\Lambda$, so this implies 
\begin{equation}
\rho_\Lambda \leq (1+z_c)^3 \rho_{M_0} \, .
\end{equation}
The detailed form of the argument gives
\begin{equation}
\rho_\Lambda \leq \frac{\pi^2}{3} (1+z_c)^3 \rho_{M_0} \simeq 410 \rho_{M_0} \, .
\end{equation}
We know in reality $\rho_\Lambda \sim 3 \rho_{M_0}$, so this bound lies within two orders of magnitude of the observed value. At this stage one can apply more detailed statistical reasoning to obtain a typical value closer to the observed value.

For this to be truly explanatory, we should envision a landscape of vacua over which the cosmological constant varies, all of which can be realized, but only a small number of which produce observers to witness them. Thus a satisfying anthropic argument for the cosmological constant requires a theoretical framework with a landscape of vacua over which the cosmological constant is finely scanned. This was famously furnished in the context of string theory by Bousso \& Polchinski \cite{Bousso:2000xa}.

One of the reasons Weinberg's anthropic argument has taken such hold is its evident success in predicting the cosmological constant, since it came well before the decisive measurement. However, it bears noting that, at the time of Weinberg's anthropic argument, Loh and Spillar \cite{Loh:1986wg} had set a limit $\rho_{\Lambda} / \rho_{M_0} = 0.1^{-0.4}_{+0.2}$ from surveys of galaxies as a function of redshift. Weinberg's assessment of this result at the time was
\begin{quote}
This is more than 3 orders of magnitude below the anthropic upper bound discussed earlier. If the effective cosmological constant is really this small, then we would have to conclude that the anthropic principle does not explain why it is so small.
\end{quote}
before going on to discuss possible problems with the experimental result. Of course, we know this bound \cite{Loh:1986wg} was off by an order of magnitude of the true value, but it is far from obvious that two orders of magnitude is better than three. 

Another potential loophole is that the anthropic bound on the cosmological constant is not a one-parameter argument. If gravitational condensation occurred at much higher redshift, the bound would be much weaker. This is possible if the amplitude of primordial density perturbations $\delta \rho / \rho \sim 10^{-5}$ were allowed to increase, which could indeed be increased by at least an order of magnitude before impacting anthropic viability, and significantly impacts the anthropic bound. Nonetheless, the apparent success of an anthropic argument for the cosmological constant sets a high bar for natural explanations, to which we now turn.

\subsection{Relaxation}

Much like the axion relaxes potentially large contributions to the $\theta$ parameter, a compelling alternative is for some dynamics to relax otherwise large contributions to the cosmological constant. 

The archetypal proposal is due to Abbott \cite{Abbott:1984qf}, which we will review in some detail here in order to set the stage for recent approaches to both the cosmological constant and electroweak hierarchy problems. Abbott's proposal introduces a new confining sector coupled to an axion-like particle with a classical shift symmetry $\varphi \rightarrow \varphi + c$ (not necessarily that of a Goldstone from a compact symmetry group) and the typical axion-like coupling
\begin{equation}
\frac{\alpha}{8 \pi} \frac{\varphi}{f_\varphi} F^a \tilde F^a \, .
\end{equation}
Non-perturbative effects give an axion potential
\begin{equation}
V_1 =  - \Lambda_\varphi^4 \cos(\varphi / f_\varphi)
\end{equation}
which breaks the classical shift symmetry to the discrete subgroup $\varphi \rightarrow \varphi + 2 \pi N f_\varphi$. In order to be relevant to the cosmological constant problem, $\Lambda_\varphi \leq 10^{-34}$ eV, but this is not so hard to engineer by virtue of dimensional transmutation; for an $SU(2)$ theory with six quarks, this amounts to $\alpha(M_{\rm Pl}) \leq 0.01$. The symmetry breaking scale is taken to be large, perhaps $f_\varphi \sim M_{\rm Pl}.$ 

In addition, a tilt is given to the cosine via a second term,
\begin{equation}
V_2 = \varepsilon \frac{\varphi}{2 \pi f_\varphi} \, ,
\end{equation}
where $\varepsilon < \Lambda_\varphi^4$. Here we have taken a linear perturbation, but various other deformations would also work, as long as they don't introduce additional minima over the field range we'll discuss. Since $\varepsilon$ breaks the discrete symmetry, its smallness can be technically natural, and all radiative corrections to $\varepsilon$ are guaranteed to be proportional to it. 

The vacuum energy density in this theory is given by
\begin{equation}
\rho_{\Lambda} = - \Lambda_\varphi^4 \cos(\varphi / f_\varphi) + \varepsilon \frac{\varphi}{2 \pi f_\varphi} + \dots
\end{equation}
with minima at $\varphi_n \approx 2 \pi n f_\varphi$ for small $\varepsilon$, and in these minima $\rho_\Lambda \approx  n \varepsilon - \Lambda_\varphi^4 + \dots$. Now by assumption, $\varepsilon < (10^{-34} \, {\rm eV})^4$, so we are guaranteed there is always a minimum where the total energy density is $\sim \varepsilon$, which we can make arbitrarily small.

To account for the cosmological constant, we must explain why the universe is in one of the states with a small cosmological constant, instead of another one. If we imagine starting at some arbitrary point on the potential with large, positive cosmological constant, we are in a de Sitter spacetime and over time $\varphi$ will evolve down the potential, decreasing the vacuum energy density at each step. Initially, when $\rho_\Lambda > M_{\rm Pl}^2 \Lambda_\varphi^2$ the barriers are irrelevant because of the non-zero Hawking temperature in de Sitter space, $T_H^2 = \frac{2}{3 \pi} \frac{\rho_\Lambda}{M_{\rm Pl}^2}$, so the field can undergo thermal fluctuations over the barriers (and instantons generating the barriers are moreover suppressed). Eventually, we will hit 
\begin{equation}
\rho_\Lambda < M_{\rm Pl}^2 \Lambda_\varphi^2 \leq (10^{-3} \, {\rm eV})^4 \, .
\end{equation}
(This is the reason for our $\Lambda_\varphi$, and hence $\varepsilon^{1/4}$, to be much smaller than $\rho_\Lambda$ --- it's not the step size that matters, but the point at which the barriers switch on.)
At this point the barriers become relevant, and field evolution proceeds via tunneling, i.e., bubble nucleation. For $\rho_\Lambda \ll M_{\rm Pl}^2 \Lambda_\varphi^2$, the tunneling rate per unit volume is
\begin{equation}
\Gamma/V \sim \Lambda_\varphi^4 e^{- \frac{3}{8} M_{\rm Pl}^4 / \rho_{\Lambda}}
\end{equation}
and eventually the evolution becomes quite slow. 

This all takes a long time, $10^{450}$ years for $\rho_\Lambda \sim M_{\rm Pl}^4$ to be reduced to the observed value. However, once we get there, we remain in a series of states with acceptable cosmological constant for a far longer time, $10^{10^{248}}$ years. Eventually we tunnel to a state with small, negative vacuum energy, but this is expected to undergo gravitational collapse and the game's over. In the meantime, we have a doubly exponentially long time in a realistic vacuum.

The problem is that the universe {\it only} contains vacuum energy. Any initial matter density is rapidly inflated away, and any matter density generated during a tunneling event is inflated away while awaiting the next transition. The last transition to the current vacuum can't reheat above $T_{\rm RH} \sim \varepsilon^{1/4}$, and even matter created from this is unlikely to be isotropic because the energy released by the tunneling event is primarily stored in the bubble wall. Even if you imagine raising the scales so that the step size is of order $\rho_\Lambda^{1/4}$, you are still impossibly far away from getting a realistic universe. 

Recently, attempts have been made to develop constructions inspired by the Abbott model that solve the reheating problem. A proof of principle was provided in \cite{Alberte:2016izw}, in which the key ingredient is a sector violating the null energy condition (NEC). The NEC violation induces an inflationary epoch followed by reheating and standard Big Bang cosmology, with symmetries restricting the cosmological constant to be the same before and after the NEC-violating phase. A related proposal \cite{Graham:2019bfu} involves a bounce following the relaxation epoch, after which the universe expands and proceeds through standard cosmological history. Although the added ingredients in both proposals are somewhat exotic, they pave the way towards potentially viable relaxation of the cosmological constant. A relaxation mechanism without NEC violation has recently been proposed \cite{Burgess:2021obw}, involving a very supersymmetric gravity sector coupled to a matter sector with non-linearly realized supersymmetry and an accidental approximate scale invariance. A related ``crunching'' mechanism for solving the cosmological constant problem was proposed in \cite{Bloch:2019bvc}, wherein regions of space with a large cosmological constant crunch shortly after inflation, whereas regions with a small cosmological constant are metastable and survive to late times.  

A very different sort of relaxation mechanism, famously proposed by Coleman \cite{Coleman:1988tj}, entails the relaxation of the cosmological constant at low energies by the effects of virtual wormholes. The essential argument is that including the effects of wormholes in the path integral leads to a doubly-exponential enhancement of the measure at the point where the total cosmological constant vanishes. This would then render a vanishing cosmological constant ``natural'' in the sense that it is a generic prediction. This argument encounters a number of significant challenges, including the inability to accommodate the small {\it but nonzero} observed value of the cosmological constant; the apparent existence of an inflationary epoch; and various problems with the prediction itself. Nonetheless, this highlights the potential phenomenological relevance of wormholes or gravitational instantons. For an excellent overview of recent progress, see \cite{Hebecker:2018ofv}.

\subsection{Symmetry}

There is an excellent symmetry for the cosmological constant problem: supersymmetry. We will have a fair bit more to say about supersymmetry when we turn to the electroweak hierarchy problem, but in some sense the true calling of supersymmetry should have been to solve the cosmological constant problem. In globally supersymmetric theories, the vacuum energy density vanishes exactly; quantum corrections cancel between bosons and fermions, while the supersymmetry-preserving minima of scalar potentials occur at $V = 0$. Of course, the lack of apparent supersymmetry below the TeV scale suggests that supersymmetry, if present at all, is spontaneously broken at a scale $\Lambda_{\rm SUSY}$. This leads to the prediction $\rho_\Lambda \sim \Lambda_{\rm SUSY}^4 \gtrsim (1 \, {\rm TeV})^4$, making supersymmetry a promising approach to the cosmological constant problem in theory but not in practice. The failure of supersymmetry to explain the cosmological constant in 3+1 dimensions was famously captured by Witten \cite{Witten:1994cga}:
\begin{quote}
Within the known structure of physics, supergravity in four dimensions leads to a dichotomy: either the symmetry is unbroken and bosons and fermions are degenerate, or the symmetry is broken and the vanishing of the cosmological constant is difficult to understand.
\end{quote}
Witten's emphasis on {\it four} dimensions was not accidental: in 2+1 dimensions the dichotomy disappears, as supersymmetry without degenerate boson and fermion masses can still explain a vanishing cosmological constant. Loosely speaking, the idea is that in 2+1 dimensions supersymmetry can control the vacuum and ensure the vanishing of the vacuum energy without controlling the spectrum of excited states. Although this possibility is extremely compelling, it remains to be realized in a form relevant to the cosmological constant in our 3+1 dimensional universe.

However, supersymmetry is not the only symmetry that might have something to say about the cosmological constant problem. Among other, more exotic, possibilities is an unusual discrete symmetry, ``$E \leftrightarrow -E$''. The idea, which originates with Linde \cite{Linde:2005ht} but was fleshed out further by Kaplan and Sundrum \cite{Kaplan:2005rr}, is to introduce parity partners of all normal fields with an opposite-sign Lagrangian density. The radiative contributions from the normal matter sector and its wrong-sign partner to the cosmological constant cancel, leaving only the bare contribution. We can think of this as arising from a $\mathbb{Z}_2$ energy-parity symmetry $P$ that {\it anticommutes} with the Hamiltonian, $\{H,P \} = 0$, so that an energy eigenstate ($H|E \rangle = E | E \rangle$) is transformed into one with opposite energy, $HP|E \rangle = - EP |E \rangle$.

The problem is that a Minkowski vacuum is evidently unstable to the pair production of positive- and negative-energy states. If the two sectors can be completely decoupled, this pair production process is suppressed and the Minkowski vacuum is effectively stable. If there is a Poincare-invariant state that is $P$ invariant, $P | 0 \rangle = |0 \rangle$, then $\langle 0 | \{ H, P \} | 0 \rangle = 2 \langle 0 | H | 0 \rangle = 0$, corresponding to vanishing cosmological constant. Although the matter action respects energy-parity, the gravitational action violates it.  Since gravity violates the parity, one might expect a gravitational contribution to the cosmological constant of order $\rho_\Lambda \sim \Lambda_{\rm grav}^4$, a scale corresponding to the cutoff of graviton momenta --- so the scale at which a quantized EFT of Einstein gravity must break down. To reproduce the observed cosmological constant, this implies $\Lambda_{\rm grav} \lesssim 2 \times 10^{-3}$ eV, or a length scale of $\sim 100$ microns, which is in tension with current short-distance tests. Nonetheless, this and related ideas have motivated recent work probing the (in)stability of theories with ghosts \cite{Gross:2020tph, Deffayet:2021nnt}.

\subsection{UV/IR Mixing}

Another possibility is that there is a breakdown in effective field theory, corresponding to some mixing between UV and IR physics. This is surprising but not unprecedented, as UV/IR mixing appears to be a feature of quantum gravity. If there is UV/IR mixing present in the theory of quantum gravity, one might hope to put it to work by inferring long-distance properties that might be felt at lower energies. The potential implications of quantum gravity for the cosmological constant problem (and particle physics more broadly) are reviewed in a pair of dedicated Snowmass white papers \cite{Berglund:2022qcc, Draper:2022pvk}. 

Encouragingly, a form of UV/IR mixing has already been used to understand an entirely different naturalness problem from the ones studied here. The static Love numbers of both spherical and spinning black holes vanish in 4d Einstein gravity, implying that all quadratic finite-size operators without time derivatives in the corresponding worldline effective field theory vanish for black holes. This is a naturalness problem \cite{Porto:2016zng} very much akin to the ones we have already encountered. Remarkably, this naturalness problem was recently solved with reference to a hidden $SL(2,\mathbb{R}) \times U(1)$ ``Love symmetry'' which mixes UV and IR modes \cite{Charalambous:2021kcz}. Although the solution is formally a demonstration of 't Hooft naturalness, its reliance on UV/IR mixing is an encouraging sign for applying similar ideas to other naturalness problems.

Various ideas about UV/IR mixing and the cosmological constant have been put forward, most notably by Banks \cite{Banks:1995uh}  and Horava \cite{Horava:1997dd}. Here we will focus on a proposal by Cohen, Kaplan, and Nelson \cite{Cohen:1998zx}, which has recently been the subject of further exploration \cite{Bramante:2019uub, Banks:2019arz, Cohen:2021zzr, Davoudiasl:2021aih, Blinov:2021fzl, Ramakrishna:2021sll}.\footnote{Special thanks to Patrick Draper, Isabel Garcia Garcia, and Matthew McCullough for extended discussion of \cite{Cohen:1998zx}, aspects of which are reflected here.} The essential idea is to leverage entropy bounds arising in a theory of quantum gravity to influence physics in the infrared.

Normally, an EFT in a box of size $L$ (an IR cutoff) with UV cutoff $\Lambda$ has extensive entropy, $S \sim L^3 \Lambda^3$. Inspired by black hole thermodynamics, Bekenstein formulated a series of conjectures about entropy in field theory \cite{Bekenstein:1973ur, Bekenstein:1974ax, Bekenstein:1980jp,Bekenstein:1993dz}, namely that the entropy in a box of volume $L^3$ only grows as the area of the box. Any EFT would violate this bound in a sufficiently large box, so if the bound is true, it implies that conventional field theories vastly over-count degrees of freedom. One way to reconcile these would be if there is a connection imposed between the UV and IR cutoffs of an EFT by requiring it to satisfy the conjectured bound. This would mean
\begin{equation}
L^3 \Lambda^3 \lesssim S_{\rm BH} = \pi L^2 M_{\rm Pl}^2 \Rightarrow L \lesssim \frac{M_{\rm Pl}^2}{\Lambda^3} \, .
\end{equation}
But a more refined condition is possible. An EFT satisfying the above bound contains many states with Schwarzschild radius larger than the box, which should probably not be described by a local QFT. We can exclude those by requiring the Schwarzschild radius of the maximum energy configuration (corresponding to an energy $L^3 \Lambda^4$) not to exceed the size of the box, i.e., 
\begin{equation}
L_s \sim \frac{L^3 \Lambda^4}{M_{\rm Pl}^2} \lesssim L \Rightarrow L \lesssim \frac{M_{\rm Pl}}{\Lambda^2} \, .
\end{equation}
This would imply that any EFT with a cutoff $\Lambda$ has a correlated IR cutoff $L$ when coupled to gravity. 

What to make of these correlated cutoffs? A conservative and relatively uncontroversial interpretation of the bound is that highly-occupied states are not well-described by quantum field theory due to strong gravitational back-reaction. A much more audacious interpretation is that $L$ and $\Lambda$ should be treated as true EFT cutoffs applicable at the few-particle level.\footnote{It bears emphasizing that a correlation between UV and IR cutoffs for few-particle states in a gravitational theory is not, on its own, a sign of UV/IR mixing. For instance, if the IR cutoff arises from a graviton mass, $L = 1/m_{\rm graviton}$, then the high-energy growth of scattering amplitudes implies a prosaic UV cutoff $\Lambda^3 \lesssim M_{\rm Pl} / L^2$ above which unitarity is violated.} Cohen, Kaplan, and Nelson's conjectured application to the cosmological constant is then as follows: if the IR cutoff of the Standard Model (and everything else) is taken to be comparable to the current horizon size, the corresponding UV cutoff is $\Lambda \sim 10^{-2.5}$ eV, surprisingly close to the observed value of the CC. Now, this is not wholly satisfying --- it is to some degree tautological, and an effective field theorist would expect to see features at the cutoff which are not (to our knowledge) apparent in Nature. In any event, it illustrates how conjectured properties of a theory of quantum gravity might be brought to bear to constrain otherwise-independent parameters of an EFT.

Recently this idea has been the subject of further exploration in various directions, e.g. \cite{Bramante:2019uub, Banks:2019arz, Cohen:2021zzr, Davoudiasl:2021aih, Blinov:2021fzl, Ramakrishna:2021sll}. An optimistic interpretation would suggest effects observable in precision measurements \cite{Bramante:2019uub, Cohen:2021zzr, Davoudiasl:2021aih}, while a more conservative interpretation \cite{Banks:2019arz, Blinov:2021fzl} leads to negligible corrections in precision measurements while retaining relevance to the cosmological constant problem by implying a thinning of field-theoretic degrees of freedom contributing to the vacuum energy density. There is considerable potential for further exploration of this direction.

\section{The Electroweak Hierarchy Problem}

We now turn to the final naturalness problem: the electroweak hierarchy problem. If we consider the Standard Model as an effective field theory up to some cutoff $\Lambda$, computing one-loop corrections from Standard Model fields to the Higgs mass gives us a famous quadratic divergence reminiscent of Weisskopf's result:
\begin{eqnarray}\label{eq:quadraticdivergence2}
\Delta m_H^2 = \frac{\Lambda^2}{16 \pi^2} \left( -6 y_t^2 + \frac{9}{4} g^2 + \frac{3}{4} g'^2 + 6 \lambda \right) \, .
\end{eqnarray}
 As we have emphasized earlier, the divergence itself is not a problem. In the Standard Model alone, the Higgs mass is merely a parameter fixed by measurement, and the above divergences are absorbed by a suitable renormalization procedure (or are absent entirely in some choices of regularization, such as dim reg). The Standard Model in isolation does not suffer from a hierarchy problem.
 
But the Standard Model is not, ultimately, in isolation. In a more complete theory with additional physical scales, the divergence in Eq.~\eqref{eq:quadraticdivergence2} is replaced by finite, calculable contributions (see, e.g.~\cite{Cohen:2019wxr} for extended discussion of this point). In this respect the divergence is merely a sign that the Higgs mass is sensitive to UV physics, a consequence of attempting to compute the Higgs mass in only part of a more complete theory. Indeed, if the Standard Model is all there is, the Higgs mass parameter is technically natural; the finite corrections proportional to e.g. the masses of Standard Model particles are all small. It is only in the presence of additional UV scales that the problem emerges. 

In discussing the electroweak hierarchy problem, it is typical to invoke $\Lambda \sim M_{\rm Pl}$ in the expectation that new physics should enter at the apparent scale of quantum gravity. In that case, Eq.~\eqref{eq:quadraticdivergence2} implies quantum corrections that are 32 orders of magnitude in excess of the doublet mass parameter $m_H^2$ inferred from the Higgs vev $v$ and the physical Higgs mass $m_h$. But a complete theory of quantum gravity remains elusive, so one might be tempted to speculate that the theory of quantum gravity `takes care of itself' at the scale $M_{\rm Pl}$ without inducing physical thresholds seen by the Higgs.\footnote{Properly speaking, proponents of this idea should then commit to demonstrating a proof of principle, as it is a highly non-trivial thing to ask from the theory of gravity. For valiant and qualitatively very different efforts in this direction, see e.g. \cite{Dubovsky:2013ira, Salvio:2014soa}.} However, in this case we should continue to extrapolate the Standard Model up to arbitrarily high energies, until the hypercharge gauge coupling hits a Landau pole around $10^{41}$ GeV. A UV completion of the Landau pole introduces a new scale playing the role of $\Lambda$, but now $\Lambda \gg M_{\rm Pl}$. Avoiding this conclusion through gauge coupling unification or a transition to conformal dynamics \cite{MarquesTavares:2013szc} induces additional physical scales that will enter into the Higgs mass. So the Standard Model is genuinely an effective field theory with cutoff $\Lambda$ whether or not one is concerned about the implications of quantum gravity. 

As we have already emphasized, the Higgs is not the only degree of freedom in the Standard Model whose mass poses a naturalness problem; it is simply the only one for which we do not yet know the answer. Thus it seems prudent to use naturalness as a strategy to guide the search for new physics. Naively asking that the corrections in Eq.~\eqref{eq:quadraticdivergence2} not exceed the inferred Higgs doublet mass parameter implies new physics should enter around $\Lambda \lesssim 500$ GeV. Of course, insofar as $\Lambda$ is merely a proxy for unknown microscopic physics, it may well be that the masses of new particles lie  within an order of magnitude of this estimate.

Although we know the scale implied by naturalness of the Higgs, we do not know the specific mechanism. More than four decades of of thinking about the electroweak hierarchy problem have generated a plethora of candidates. Here we will briefly review some of the canonical approaches before focusing our attention on recent developments.

\subsection{Canonical approaches}

The first thing one is tempted to do when confronted by the hierarchy problem is to erase the apparent hierarchy itself, bringing down the cutoff of the Higgs sector or the entire Standard Model. Indeed, this was the nature of the first attempted solution to the hierarchy problem, {\it technicolor} (due to Weinberg \cite{Weinberg:1975gm} and Susskind \cite{Susskind:1978ms}), which attempted to replicate the success of the proton mass prediction by imagining that electroweak symmetry was broken by the vacuum condensate of a strongly coupled group. The five-dimensional holographic duals of technicolor are Randall-Sundrum models \cite{Randall:1999ee, Randall:1999vf}. In these cases, the Higgs is not an elementary degree of freedom, and the cutoff is provided by compositeness of the Higgs itself. Alternately, we could imagine leaving the Higgs alone and lowering the scale of quantum gravity, so that all field theoretic physics reaches an end at the cutoff. This is the nature of solutions such as large extra dimensions \cite{ArkaniHamed:1998rs, Antoniadis:1998ig}. More recently, a third extra-dimensional option has come to the fore, where the geometry is set by a five-dimensional dilaton whose background profile varies linearly in the extra dimension \cite{Baryakhtar:2012wj, Giudice:2016yja, Giudice:2017fmj}. Recent progress on warped compactifications relevant to naturalness is summarized in a dedicated Snowmass white paper \cite{Agrawal:2022rqd}. 

The problem with pure lowered-cutoff solutions is that they generically do not predict any separation between the Higgs and the scale of new physics. That is, the typical expectation of the Higgs mass is of order $m_H^2 = c \Lambda^2$, with $c\sim \mathcal{O}(1)$. Such theories then predict a host of particles close in mass to the Higgs, as well as a host of higher-dimensional operators suppressed by a low cutoff. The non-observation of new particles close in mass to the Higgs, as well as strong bounds on irrelevant operators correcting the Standard Model, suggests that this mechanism is not operative on its own. But if the proton mass is not a viable analogy (to our knowledge) for the naturalness of the Higgs mass, it is sensible to consider whether the other mechanisms realized by Nature might play a role. 

\subsubsection*{Supersymmetry}

As we have seen, the electron mass was rendered 't Hooft natural by a chiral symmetry. A scalar enjoys no such protection on its own, but could `borrow' the chiral symmetry of a fermion if there is an additional symmetry relating bosons and fermions. This is the sense in which {\it supersymmetry} can solve the electroweak hierarchy problem, by making the mass of a scalar proportional to that of a fermion, which is itself protected by chiral symmetry. This symmetry must be softly broken in order to be consistent with the non-observation of degenerate superpartners. For an excellent review, see e.g.~\cite{Martin:1997ns}.

One of the conceptual virtues of supersymmetry is that it provides a very concrete, calculable realization of the expectation we have attached to Eq.~\eqref{eq:quadraticdivergence2}, namely that the quadratic divergence is merely a proxy for finite, calculable contributions in a more complete microscopic theory. In supersymmetric extensions of the Standard Model, the quadratic divergences indeed vanish and are replaced by the mass splittings between Standard Model particles and their superpartners.\footnote{Corrections to the Higgs mass parameter remain logarithmically divergent unless the detailed mechanism of supersymmetry breaking is specified, in which case they can be rendered finite.} 

Of course, after decades of mounting expectations for the appearance of supersymmetry at the TeV scale, the LHC has found no evidence for superpartners. At this point it is fair to say that supersymmetry did not appear where naturalness arguments led us to expect it.\footnote{There are many lessons to be learned from this failure. It should be cause for considerable introspection.} That is not to say that supersymmetry may not appear somewhere above the TeV scale, but this leaves a fair bit of daylight between Nature and naturalness expectations. The degree to which naturalness expectations are violated is often quantified by fine-tuning using various measures (e.g. the Barbieri-Giudice measure, \cite{Barbieri:1987fn}), but ultimately we do not know how Nature computes fine-tuning, making it difficult to draw quantitative conclusions.

\subsubsection*{Compositeness}

The other path already chosen by Nature is the combination of a spontaneously broken global symmetry and compositeness, as realized by the charged and neutral pions. Such a {\it composite Higgs} \cite{Kaplan:1983fs} is naturally separated from the scale of compositeness itself by a moderate amount, depending on the degree to which Standard Model couplings violate the global symmetry protecting the Higgs. This generically predicts the modification of Higgs couplings relative to Standard Model expectations \cite{Giudice:2007fh}, as well as new degrees of freedom around the TeV scale. For an excellent review, see \cite{Panico:2015jxa}. There is still room for a composite Higgs to satisfy generic naturalness expectations, but the tension will increase significantly if the LHC finds no evidence for Higgs coupling deviations or new particles beneath a few TeV.  Recently, considerable progress has been made in alleviating the tuning in composite Higgs models that is otherwise implied by the absence of large Higgs coupling deviations \cite{Csaki:2017cep, Durieux:2021riy}. 

\subsubsection*{Anthropics}

A final ``conventional'' possibility is anthropics: nothing protects the Higgs mass, but rather there are many vacua of the Standard Model over which the Higgs mass varies according to some statistical distribution. If there is then a mechanism for selecting from the tail of the distribution with smaller Higgs masses, one has an explanation for the observed Higgs mass that does not rely on symmetries or a low cutoff, much like the proposed anthropic explanation of the cosmological constant. 

The prevailing version of this argument uses anthropic pressure to understand the weak scale in a universe where the dimensionful parameters of the Standard Model (i.e., the Higgs mass, or equivalently the vacuum expectation value $v$) vary, but the dimensionless quantities are held fixed. In this case, $v$ is bounded from above to be near its observed value by an argument known as the Atomic Principle \cite{Agrawal:1997gf}. Recall that for $v = v_{SM}$ the lightest baryons are the proton and neutron, of which the proton is lighter because the splitting due to quark masses exceeds the electromagnetic energy splitting: $m_n - m_p = (3 v/v_{SM} - 1.7)\, {\rm MeV}$. Free neutrons decay into protons, with a reaction energy $Q = m_n - m_p - m_e = (2.5 v/v_{SM} - 1.7)\, {\rm MeV}.$

But in nuclei there is a binding energy that stabilizes the nuclei. Without going into the details, it suffices to note that the long-range part of the nucleon-nucleon potential is due to single pion exchange, with a range of $\sim 1/m_\pi$. For small $u, d$ masses $m_\pi \propto ((m_u + m_d) f_\pi)^{1/2}$, so (neglecting the weak dependence of $\Lambda_{QCD}$ on $v$) we have $m_\pi \sim v^{1/2}$. Mocking up the binding energy in deuterons (the most weakly bound system) as a square well with a hard core to mimic short-range repulsion gives
$$B_d \simeq \left[2.2 - 5.5 \left(\frac{v - v_{SM}}{v_{SM}} \right) \right] \, {\rm MeV}$$
for small $v-v_{SM}$.

Now we see that as we increase $v$, we will eventually reach the point where $B_d < Q$ and the neutron is no longer stabilized by nuclear binding energy. This occurs for $v/v_{SM} \gtrsim 1.2$, which is a tight bound, indeed! The deuteron is fairly important, since all primordial and stellar nucleosynthesis begins with deuterium. But this is not an airtight bound, as nuclei could form in violent astrophysical processes. The binding energies for heavier nuclei are larger, but for $v / v_{SM} \gtrsim 5$
typical nuclei no longer stabilize the neutron against decay. 

Assuming that stable protons and complex atoms are required for observers to form, this provides an anthropic pressure that favors $v \lesssim v_{SM}$. But it is clear that a robust constraint only exists if dimensionless couplings are held fixed; variation of the yukawas allows these constraints to be naturally evaded (although other catastrophic boundaries may be encountered, see e.g.~\cite{Hall:2014dfa}). Indeed, it is possible to imagine a ``weak-less'' universe where the gauge group of the Standard Model is $SU(3)_c \times U(1)_{\rm em}$, and fermions appear in vector-like representations \cite{Harnik:2006vj}. It has been argued that such a universe undergoes big-bang nucleosynthesis, matter domination, structure formation, and star formation --- i.e., sufficient stages of development to produce some form of observers. Of course, truly demonstrating that such a theory is capable of reproducing the physics necessary for forming observers is challenging, but suffices to indicate that anthropic reasoning applied to the weak scale is sufficiently permeable. 

Given the appeal of an anthropic explanation for the cosmological constant problem, it is sensible to consider whether this informs a possible anthropic explanation for the weak scale. In general, a landscape providing an anthropic explanation for both problems must contain enough vacua to scan both the cosmological constant and the weak scale with sufficient precision. This is a significant demand on top of the vacuum multiplicity required to scan the cosmological constant alone, making it seem more economical to set the weak scale naturally even if the cosmological constant is set anthropically. However, it may be possible to correlate the value of the cosmological constant with the weak scale \cite{Arvanitaki:2016xds, Arkani-Hamed:2020yna} in such a way that the landscape need only contain sufficient vacua to scan the former and not the latter.

\subsection{Recent developments}

Thus far, there is no experimental evidence for compositeness or supersymmetry as a solution to the electroweak hierarchy problem. But as we have emphasized, the fact that Nature does not realize a specific mechanism for naturalness of the electroweak scale does not mean that the electroweak scale is unnatural, or that naturalness itself has failed. Perhaps what is called for are new ideas, ones that do not necessarily replicate one of the mechanisms already used by Nature. In what follows we will survey some of the new directions that have been developed since the last Snowmass process.

\subsubsection*{Discrete symmetries}

One interesting direction is to retain the symmetry-based approach but expand the scope of possible symmetries. The most apparent possibility is to work with discrete symmetries, rather than continuous ones. The appeal is that the new particles required by a discrete symmetry need not carry the same Standard Model quantum numbers, and so are less strongly constrained by data from the LHC. 

There are by now many different examples of ``neutral naturalness'' \cite{Craig:2014aea}, but the simplest is the original: the Twin Higgs \cite{Chacko:2005pe}. The idea is to introduce a mirror copy of the Standard Model along with a $\mathbb{Z}_2$ symmetry exchanging each field with its mirror counterpart. On top of this, one needs to assume an approximate global symmetry in the Higgs sector; this global symmetry need not be exact, and is violated by all SM yukawa and gauge couplings, but should be an approximate symmetry of the Higgs potential. This does not stabilize the Higgs mass to arbitrarily high scales \cite{Contino:2017moj}, but rather postpones the scale at which true solutions to the hierarchy problem (such as supersymmetry \cite{Falkowski:2006qq, Chang:2006ra, Craig:2013fga} or compositeness \cite{Chacko:2005vw, Geller:2014kta, Barbieri:2015lqa, Low:2015nqa}) must appear. Experimental signatures of neutral naturalness are quite different from those typically associated with supersymmetry or compositeness, insofar as the partner particles predicted by the discrete symmetry populate a Hidden Valley \cite{Strassler:2006im}. The most promising signals depend on the nature and quality of the discrete symmetry, but typically include Higgs coupling deviations and Higgs decays into invisible or long-lived particles \cite{Craig:2015pha, Chacko:2015fbc}. Higgs coupling deviations generically provide the strongest constraint on the natural parameter space of these models, though tuning and Higgs couplings may be decoupled \cite{Durieux:2022sgm}. Light, stable partner particles can be copiously produced in the early universe and give rise to promising signatures in the CMB and large scale structure \cite{Craig:2016lyx, Chacko:2016hvu, Chacko:2018vss}, exemplifying the growing relevance of cosmological observations to the electroweak hierarchy problem.

Just as the Twin Higgs leverages a discrete symmetry to produce an accidental global symmetry, folded supersymmetry \cite{Burdman:2006tz, Cohen:2015gaa} leverages a discrete symmetry to produce an accidental supersymmetry. Theories with completely neutral scalar partners for Standard Model fermions require an interplay between a discrete symmetry, a continuous global symmetry, and supersymmetry, and have only been discovered more recently \cite{Cohen:2018mgv, Cheng:2018gvu}.

Models and signatures of neutral naturalness have been explored extensively in the past decade. There is a comprehensive Snowmass white paper dedicated to neutral naturalness \cite{Batell:2022pzc}, to which we refer the reader for further details. Some cosmological implications of neutral naturalness are also discussed in Snowmass white papers on light cosmological relics \cite{Dvorkin:2022jyg} and early-universe model building \cite{Asadi:2022njl}. 

The relevance of discrete symmetries to the hierarchy problem extends well beyond the framework of neutral naturalness. Nonlinearly realized discrete symmetries can significantly alter naive expectations of naturalness and provide new approaches to the hierarchy problem \cite{Hook:2018jle, Das:2020arz}.

\subsubsection*{Relaxation}

In some sense, neutral naturalness is a conservative ``new'' idea for the electroweak hierarchy problem, in that it retains a familiar mechanism (symmetry protection) while pushing the specific realization in a new direction. But the past decade has seen the emergence of several entirely new approaches, in some cases inspired by proposals for the strong CP and cosmological constant problems. Chief among these is relaxation of the weak scale. Aspects of these approaches are also discussed in a Snowmass white paper on early-universe model building \cite{Asadi:2022njl}.

The original incarnation is the {\it relaxion} \cite{Graham:2015cka}, inspired by the Abbott model for the cosmological constant, featuring a QCD axion-like particle $\phi$ coupled to the Standard Model with an additional inflationary sector whose properties are necessarily somewhat special.  The simplest realization involves enlarging the Standard Model with the following terms:
\begin{equation}
\delta \mathcal{L} = (-M^2 + g \phi) |H|^2 + V(g \phi) + \frac{\alpha_s}{8 \pi} \frac{\phi}{f} \tilde G^{\mu \nu} G_{\mu \nu} \, ,
\end{equation}
where $M$ is of the order of the cutoff of the SM Higgs sector, $H$ is the Higgs doublet, $g$ is the dimensionful coupling that breaks the shift symmetry, and $V(g\phi) \sim g M^2 \phi + g^2 \phi^2 + \dots$
 parameterizes the non-derivative terms solely involving $\phi$. We will be interested in field values of $\phi$ that greatly exceed $f$, so we should understand it as a non-compact field (or a compact field imbued with an effective period much longer than $2 \pi f$). When $g/M \rightarrow 0$ the Lagrangian has a shift symmetry $\phi \rightarrow \phi + 2 \pi f$, and $g$ can be treated as a spurion for breaking of the shift symmetry. 

Below the QCD confinement scale, the coupling between $\phi$ and the gluon field strength gives rise to the familiar periodic axion potential
\begin{equation}
\frac{\alpha_s}{8 \pi} \frac{\phi}{f} \tilde G^{\mu \nu} G_{\mu \nu} \rightarrow \Lambda_{\rm QCD}^4 \cos(\phi/f) \, .
\end{equation}
For values of the Higgs vev near the Standard Model value, the height of the cosine potential is
\begin{equation}
\Lambda^4 \sim f_\pi^2 m_\pi^2 \sim yv f_\pi^3 \, ,
\end{equation}
where $m_\pi^2$ changes linearly with the quark masses, and so the barrier height is linearly proportional to the Higgs vev (at least roughly speaking; there are of course logarithmic corrections from the contributions to QCD running).

Now the idea is clear: starting at values of $\phi$ such that the total Higgs mass is large and positive, and assuming the slope of the $\phi$ potential causes it to evolve in a direction that lowers the Higgs mass, the $\phi$ potential will initially be completely dominated by the $g \phi$ potential terms, until the point at which the total Higgs mass-squared goes from positive to negative and the Higgs acquires a vacuum expectation value. At this point the wiggles due to the quark masses grow linearly in the Higgs vev, and generically $\phi$ will stop when the slope of the QCD-induced wiggles matches the slope of $V(\phi)$. This classical stopping point occurs when the maximum slope of the cosine potential is of the same order as the linear tilt,
\begin{equation}
g \sim \frac{y v f_\pi^3}{M^2 f} \, .
\end{equation}
This allows for a light Higgs (i.e., a small total Higgs mass-squared and small electroweak scale) relative to a cutoff $M$ provided $g/M \ll 1$. For example, with a QCD axion decay constant $f = 10^9$ GeV and $M \sim 10^7$ GeV we have $g/M \sim 10^{-30}$. 

So far we have only accounted for the parametrics of the potential, neglecting the actual dynamical process. In the minimal realization of the relaxion mechanism, $\phi$ is made to roll slowly by imagining that its evolution occurs during a period of inflation, such that Hubble friction provides efficient dissipation of kinetic energy in $\phi$. Combining all constraints, this simplest model does not stabilize the weak scale all the way to $M_{\rm Pl}$; the cutoff of the theory is at most
\begin{equation}
M \lesssim \left(\frac{\Lambda_{\rm QCD}^4 M_{\rm Pl}^3}{f} \right)^{1/6} \sim 10^7 \; {\rm GeV} \times \left( \frac{10^9 \; {\rm GeV}}{f} \right)^{1/6} \, .
\end{equation}

Unfortunately, even if all of these criteria are satisfied, there is an observational problem with this simplest scenario. The field $\phi$ stops not at the minimum of the QCD cosine potential (for which the effective $\theta$ angle is zero), but is rather displaced by an amount proportional to the slope of $\phi$. This amounts to $\theta \sim 1$, which is excluded (as we have seen) by bounds on hadronic EDMs. So the minimal mechanism is ruled out by a natural prediction, though it is certainly no fault of the mechanism itself. This can be ameliorated without extra ingredients by coupling the relaxion to the inflaton in such a way that the slope of $\phi$ decreases after inflation, reducing the contribution to $\theta$. This has the effect of lowering the scale at which the model must be UV completed, leading to $M \lesssim 30$ TeV for $\theta \lesssim 10^{-10}$. The most striking experimental signatures in minimal relaxion scenarios involve the relaxion-Higgs mixing induced by the cosine potential, leading to signals associated with a new, light Higgs-like scalar \cite{Flacke:2016szy}. A simple variation without stringent constraints from hadronic EDMs entails repeating the same ingredients, but the relaxion is instead the axion of another gauge group for which constraints on the $\theta$ parameter are weaker or nonexistent. This scenario should involve quarks of a new gauge group that are also charged under the electroweak gauge group, with attendant Hidden Valley experimental signatures \cite{Beauchesne:2017ukw}. For further development of the relaxion paradigm, see e.g. \cite{Espinosa:2015eda,Hardy:2015laa, Batell:2015fma, Choi:2015fiu, Hook:2016mqo, Nelson:2017cfv, Davidi:2018sii}.

Many of the open questions about the relaxion involve physics in the UV. There are challenges to protecting the shift symmetry of the relaxion over the vastly trans-Planckian excursions in field space required to explain the value of the weak scale, as enumerated by the Swampland program; for recent summaries of relevant considerations, see e.g.~\cite{Harlow:2022gzl, Draper:2022pvk}. One possibility is to accumulate effectively trans-Planckian flat potentials via axion monodromy \cite{Silverstein:2008sg, Ibanez:2015fcv} or clockwork \cite{Kaplan:2015fuy}. The advent of the relaxion catalyzed the development of other cosmological approaches to the electroweak hierarchy problem, for the most part leveraging vacuum selection during an inflationary epoch to preferentially populate a universe with the observed value of the weak scale \cite{Geller:2018xvz, Cheung:2018xnu, Csaki:2020zqz, TitoDAgnolo:2021nhd, TitoDAgnolo:2021pjo}.

\subsubsection*{Reheating}

An alternative that proceeds from similar inspiration is to put many copies of the Standard Model in the same universe, but explain why one copy acquires the dominant energy density \cite{Arkani-Hamed:2016rle}. This proposal, known as $N$-Naturalness, is briefly reviewed in a Snowmass white paper on light cosmological relics \cite{Dvorkin:2022jyg}; here we summarize some key features.

The idea is to envision $N$ sectors which are mutually decoupled. For simplicity, we could take it to be $N$ copies of the Standard Model, though this is not an important restriction. From copy to copy, we imagine the Higgs mass parameters are distributed in some range from $-\Lambda_H^2$ to $\Lambda_H^2$ according to some probability distribution. For a wide range of distributions, the generic expectation is that some sectors have accidentally small Higgs masses, $m_H^2 \sim \Lambda_H^2 / N$. For large enough $N$, this implies that there is a sector whose electroweak scale is well below the cutoff, which we might identify with ``our'' Standard Model. Reversing the argument, this implies that the cutoff of the theory should be $\Lambda_H \sim \sqrt{N} |m_H|$. For example, a cutoff of 10 TeV corresponds to $N= 10^4$, whereas a cutoff of $10^{10}$ GeV requires $N = 10^{16}$. 

There is another factor in play when $N$ is large. While the naive scale of quantum gravity is $M_{\rm Pl}$, in the presence of a large number of species the scale at which gravity becomes strongly coupled is lowered, $\Lambda_G^2 \sim M_{\rm Pl}^2 / N.$ This implies the effective Planck scale should be at least $M_{\rm Pl}^2 \sim N \Lambda_G^2$. Solving the entire hierarchy problem this way would entail $N = 10^{32}$. However, this lowers the cutoff of quantum gravity to the weak scale, and gives us the usual problems associated with a low cutoff. But we would naturally have one sector with the observed value of the weak scale and a Higgs cutoff associated with the cutoff of quantum gravity for $N= 10^{16}$, for which $\Lambda_H = \Lambda_G = 10^{10}$ GeV. Alternately, we could preserve a notion of grand unification for $N = 10^{4}$, for which quantum gravity grows strong at $10^{16}$ GeV, and something like supersymmetry enters at $\Lambda_H = 10$ TeV to cut off the Higgs sector. 

The question, then, is to explain why this sector with ``our'' Standard Model is populated, while all of the other sectors are not. As with the relaxion, this is accomplished through cosmology. In a universe with many sectors, the universe is populated by whatever sectors are abundant. If all sectors had a thermal abundance, there would be an enormous contribution to the energy density of the universe, and we would not have any ability to understand why we are the sector with the smallest scales. Thus we can imagine a cosmological mechanism that preferentially reheats sectors with smaller scales. The simplest way to accomplish this is to imagine an inflationary epoch, followed by reheating due to the decay of some reheaton. To avoid tuning, this reheaton should couple universally to all sectors. The Standard Model can be preferentially reheated (i.e., absorb most of the energy from the reheaton decays) if the branching ratio of the reheaton to each sector scales like an inverse power of the (absolute value of the) Higgs mass-squared in each sector. 

Remarkably, this is precisely what happens if the reheaton is lighter than the Higgs in each sector --- a parametric but technically natural requirement of the theory. Assuming this is the case, the reheaton predominantly decays into fermions of sectors where electroweak symmetry is broken, whereas when electroweak symmetry is unbroken the dominant decay is into gauge bosons. Thus the decay rate into broken-phase sectors scales as $1/m_h^2$, while the decay into unbroken-phase sectors scales as $1/m_H^4$. Reheaton decays therefore prefer a sector with broken electroweak symmetry and the smallest possible value of $m_h$.  The resulting energy density of each sector is proportional to the decay width,
\begin{equation}
\frac{\rho_i}{\rho_{us}} \simeq \frac{\Gamma_i}{\Gamma_{us}} \, .
\end{equation}
This leads to some energy density in the sectors nearest to ours in mass, with attendant predictions for dark radiation within the reach of future CMB experiments \cite{Choi:2018gho}.

\subsubsection*{UV/IR mixing}

One way to frame the hierarchy problem is as a separation of UV physics from IR physics in effective field theory: the theory in the far UV knows nothing about the theory in the far IR, and cannot generically produce IR scales well-separated from the fundamental UV scale (with the exception of special mechanisms, such as dimensional transmutation, that we have encountered earlier). From this perspective, a new approach to the hierarchy problem might entail linking the far UV and the far IR.

What are the prospects of UV/IR mixing for the hierarchy problem? As we have already seen in our discussion of the cosmological constant problem, we might expect a theory of quantum gravity to feature UV/IR mixing. Whether this UV/IR mixing has any relevance to the weak scale is an open question, but there are a number of promising possibilities. These are summarized quite comprehensively in a dedicated Snowmass white paper \cite{Draper:2022pvk}, and so our discussion here will remain fairly concise.

Some of the most promising opportunities arise in the context of the Swampland program, which articulates conjectured constraints on effective field theories from consistent embedding in a theory of quantum gravity. Perhaps the most famous among the Swampland conjectures is the Weak Gravity Conjecture (WGC) \cite{ArkaniHamed:2006dz}, which formalizes the sense in which ``gravity is the weakest force.''  For a comprehensive review of Swampland conjectures, see \cite{Palti:2019pca}; for a review focused on the Weak Gravity Conjecture and its relatives, see \cite{Harlow:2022gzl}. The possible relevance of these conjectures to the electroweak hierarchy problem is illustrated by a proposal first made by Cheung \& Remmen \cite{Cheung:2014vva} to use the (electric) Weak Gravity Conjecture to bound the weak scale.

In its simplest form, the WGC posits that an abelian gauge theory coupled to gravity must contain a state of charge $q$ and mass $m$ satisfying
\begin{equation}
q g > \frac{m}{M_{\rm Pl}}
\end{equation}
which amounts to the statement that gravity is the weakest force, since this implies the gauge force between two charges exceeds the gravitational one. Cheung \& Remmen noted that writing the inequality as $m < q g M_{\rm Pl}$ had the effect of bounding a possibly UV-sensitive parameter (the mass $m$, potentially additively sensitive to short-distance physics) by a UV-insensitive one (the coupling $g$, which is only logarithmically sensitive to short-distance physics). This could be applied to the electroweak hierarchy problem by extending the Standard Model to include an unbroken $U(1)$ and some particle charged under it whose mass satisfies the WGC and is controlled by electroweak symmetry breaking. A natural candidate is gauging $U(1)_{B-L}$, which can be rendered anomaly-free by adding a right-handed neutrino $\nu_R$. Current bounds on $U(1)_{B-L}$ require $qg \lesssim 10^{-24}$. 

In this case neutrino masses arise from a yukawa coupling to the Higgs, giving Dirac neutrino masses of the form
\begin{equation}
y_\nu v \bar \nu_L \nu_R + {\rm h.c.}
\end{equation}
The lightest neutrino has the largest charge-to-mass ratio, and if there are no other light particles in the spectrum it is the natural candidate to satisfy the WGC. For a neutrino mass around $m_\nu \sim 0.1$ eV,  
if $qg \sim \frac{m_\nu}{M_{\rm Pl}} \sim 10^{-29} $ (consistent with current bounds) then the WGC is just barely satisfied. If the values of the yukawa coupling $y_\nu$ and $U(1)_{B-L}$ coupling $qg$ are held fixed, then higher values of the Higgs vev $v$ would violate the WGC. One could then imagine that consistency of quantum gravity places an upper bound on $v$.

Of course, there are many ways in which this argument could fail: there could be lighter states charged under $U(1)_{B-L}$ that satisfy the WGC; the WGC could be satisfied in the underlying theory by varying $y_\nu$ and $qg$; etc. Unfortunately, even taking the premises to be true, the argument itself fails due to a related conjecture. The magnetic form of the WGC posits that the cutoff $\Lambda$ of a purely electric description of an Abelian gauge theory with charged states must satisfy $\Lambda \lesssim q M_{\rm Pl}$
where here the cutoff could correspond to e.g.~the scale of monopoles in the theory or some other breakdown of the purely electric description. This would imply the above construction breaks down at the scale of neutrino masses, and additional degrees of freedom associated with $\Lambda$ would appear well before the scale $v$. The proposal can be revived by considering a distinct $U(1)$ whose charged states lie closer to the weak scale and acquire some mass from the Higgs \cite{Craig:2019fdy}, in which case the bound from the magnetic WGC rises above the weak scale. This presents a number of novel experimental signatures, including new states around the TeV scale coupled to the Higgs boson and, potentially, an extremely weak long-range force acting on dark matter. Although there are various possible caveats and potential loopholes \cite{Craig:2019fdy, Harlow:2022gzl}, the proposal illustrates a sense in which Swampland conjectures may be relevant to the electroweak hierarchy problem. Indeed, there are a number of related ways that Swampland conjectures may be brought to bear to explain the value of the weak scale \cite{Ibanez:2017oqr, Montero:2019ekk, March-Russell:2020lkq}.

The Weak Gravity Conjecture and other Swampland conjectures amount to a sort of ``implicit'' UV/IR mixing, in which the parameter space of an EFT is bounded by generic criteria without reference to the microscopic physics responsible. There are also examples of theories exhibiting various forms of ``explicit'' UV/IR mixing. A very concrete example with immediate relevance to the hierarchy problem involves worldsheet modular invariance in non-supersymmetric string theory \cite{Dienes:2001se, Abel:2021tyt}. Other examples include the vanishing black hole Love numbers mentioned earlier, as well as non-commutative field theories \cite{Minwalla:1999px, Craig:2019zbn}, field theories with subsystem global symmetries \cite{Gorantla:2021bda}, and certain non-integrable quantum field theories in two dimensions \cite{Dubovsky:2013ira}. The relevance of these latter examples to the electroweak hierarchy problem is less apparent, but the exploration of theories featuring UV/IR mixing is likely to bear further fruit. At the very least, it promises to reveal new phenomena in quantum field theory.

\subsubsection*{Self-organized criticality}

There is one distinguished value for the mass-squared parameter of the Higgs doublet: $m_H^2 = 0$. For $m_H^2 < 0$ electroweak symmetry is spontaneously broken, while for $m_H^2 > 0$ it is preserved, rendering $m_H^2 = 0$ the critical value (at zero temperature) separating the two phases of electroweak symmetry. The fact that $|m_H^2 / M_{\rm Pl}^2| \ll 1$ (for, say, $\Lambda \sim M_{\rm Pl}$) amounts to the statement that we are surprisingly close to the critical point. As there are systems that drive themselves to their critical points --- a phenomenon known as self-organized criticality \cite{Bak:1987xua} --- it is inviting to consider whether something along these lines might solve the electroweak hierarchy problem.\footnote{To my knowledge, the first suggestion that self-organized criticality might be relevant to the electroweak hierarchy problem was made (ironically) by David B.~Kaplan in his 1997 TASI lectures \cite{SOC}, while a more earnest suggestion appears in \cite{Giudice:2008bi}.} 

Naively, it is difficult to realize self-organized criticality in Lorentz-invariant quantum field theories, since most instances involve both driving and dissipation. Nonetheless, recent years have seen several concrete proposals for something like self-organized criticality as an explanation of the weak scale. The first of these \cite{Eroncel:2018dkg} involves the interplay between the Higgs field and a modulus field in a 5d Randall-Sundrum model, with the Higgs instability being connected via the modulus field to violation of the Breitenlohner-Freedman bound far from the UV boundary.\footnote{For related work, see \cite{Pomarol:2019aae}.} Subsequently, analogs of self-organized criticality were developed in a cosmological setting \cite{Khoury:2019ajl,Kartvelishvili:2020thd,Giudice:2021viw}, in which fluctuations of scalar fields during an inflationary epoch lead to localization close to a critical point. Once again, new experimental signatures arise in connection with the Higgs. The simplest application of self-organized localisation \cite{Giudice:2021viw} to the electroweak hierarchy problem, for example, requires modifying the running of the Higgs self-coupling via new states near the weak scale. More broadly, these concrete examples are an encouraging indication of the prospects for self-organized criticality in understanding the electroweak hierarchy problem.

\section{Looking Forward}

With that, our journey through the main outstanding naturalness problems of high-energy physics (and their recently proposed solutions) comes to an end. We have seen something of the problems themselves, their historical solutions, and the proliferation of new approaches that have emerged in the course of the past decade. Among other things, these new approaches are distinguished by the novelty of their experimental signatures, bringing entirely new observables and experiments to bear in the search for signs of naturalness.

 These approaches also underline the extent to which naturalness problems are connected. Although the cosmological constant problem, the electroweak hierarchy problem, and the strong CP problem vary widely in both dimensionality and severity, their proposed solutions have much in common; see, for instance, the table below. If nothing else, this is a reminder that something can be gained from thinking of these naturalness problems together, rather in isolation. It may well be that a recently-proposed solution proves most fruitful when applied to a different problem from the one it was invented to address. \\

\begin{table}[h]
\begin{center}
\begin{tabular}{|c||c|c|c|}
\hline
& Strong CP Problem & CC Problem & Hierarchy Problem\\ \hline \hline
Cts. symmetry &  $U(1)_{PQ}$ & SUSY & SUSY, global \\ \hline
Disc. symmetry & P/CP & $E \rightarrow -E$ & $\mathbb{Z}_2$ \\ \hline
Relaxation & Axion & Abbott & Relaxion \\ \hline
Anthropics & Tied to CC? & Structure formation & Atomic principle \\ \hline
UV/IR mixing & ? & Holography & WGC/NCQFT/... \\ \hline
\end{tabular}
\end{center}
\label{tab:solutions}
\end{table}%
What lies ahead? A number of the new paths sketched here are in the earliest stages of exploration, and will doubtlessly develop further in the coming years. Further exploration of UV/IR mixing seems particularly promising, at the very least because it remains a relatively unexplored facet of quantum field theory and gravity with transformative potential. Although there are not, at present, any completely satisfying applications of UV/IR mixing to the marquee naturalness problems of high-energy physics, it would be premature to conclude that ``there is no there there.'' Before discovery there is always exploration, and the motivation for exploring UV/IR mixing with an eye towards naturalness problems is abundant. Self-ordered criticality is also quite promising in this regard; now that there is proof of principle in relativistic settings, there is considerable room for further exploration.

There are also numerous developments in adjacent subfields of high-energy theory that have yet to be applied directly to naturalness problems, but seem destined to play a role. The amplitudes program is perhaps the most striking example, as it has recently provided abundant evidence that the renormalization of irrelevant operators in effective field theories enjoys surprising properties that motivate some refinement of naturalness expectations. For instance, the unexpected zeroes in the one-loop dimension-6 matrix of anomalous dimensions in the Standard Model EFT is best understood from helicity selection rules \cite{Cheung:2015aba}; analogous surprises persist even at two loops \cite{Bern:2020ikv}. Such surprising zeroes extend to Wilson coefficients of irrelevant operators as well \cite{Arkani-Hamed:2021xlp}, which can be understood at least in part using on-shell techniques \cite{Craig:2021ksw, Rose:2022njd}. Of course, there may be limits to how much we can learn about the naturalness problems of marginal and relevant operators from an improved understanding of irrelevant ones, but these examples suggest that naturalness expectations should be treated with care.

More broadly, our understanding of quantum field theory is far from static, and in particular the understanding of symmetries has evolved considerably since the articulation of 't Hooft naturalness. The ordinary symmetries typically applied to naturalness problems have subsequently been joined by a plethora of generalizations, including higher-form symmetries \cite{Gaiotto:2014kfa}, higher-group symmetries \cite{Kapustin:2013uxa, Kapustin:2014zva}, subsystem symmetries \cite{Seiberg:2019vrp}, and non-invertible symmetries \cite{Bhardwaj:2017xup}. It seems quite likely that at least some of these generalized symmetries can be brought to bear on familiar naturalness problems. This is far from wishful thinking. For instance, higher-form symmetries already imbue the masslessness of the photon with a genuine notion of 't Hooft naturalness by making it the goldstone boson of a spontaneously-broken one-form global symmetry, while higher-group symmetries have given rise to new constraints on the phenomenology of axion-Yang-Mills theories \cite{Brennan:2020ehu}. If the next decade sees the emergence of a genuinely new and compelling approach to naturalness problems leveraging symmetries, it is likely to come from this direction.

Time will tell if more refined views as to the uniformity of Nature would have been useful to the particle theorist. As ever, we must look to experiment for ultimate guidance. But in the meantime, the motivation for thinking about naturalness remains strong. There are many paths to be followed, and yet more paths to be discovered.

\section*{Acknowledgments}

It is a pleasure to thank innumerable colleagues over the years for helpful conversations about naturalness that have informed this white paper, including Prateek Agrawal, Nima Arkani-Hamed, Tim Cohen, Csaba Csaki, Savas Dimopoulos, Michael Dine, Patrick Draper, Isabel Garcia Garcia, Gian Giudice, Dan Green, Howie Haber, Simon Knapen, Seth Koren, John March-Russell, Matthew McCullough, Ann Nelson, Michael Peskin, Matt Reece, Grant Remmen, Matt Strassler, Raman Sundrum, Dave Sutherland, Arkady Vainshtein, and Zhengkang Zhang. I am indebted to Tim Cohen, Patrick Draper, Isabel Garcia Garcia, Matthew McCullough, and Raman Sundrum for useful comments on a draft of this white paper. Special thanks to Patrick Draper and Ira Rothstein for both the impetus to write this white paper and generosity with respect to Snowmass deadlines.  This work was supported in part by the U.S. Department of Energy under the grant DE-SC0011702.

\bibliographystyle{JHEP}
\bibliography{natrefs}

\end{document}